
\def\sm{standard model}
\def\susy{supersymmetry}
\def\susic{supersymmetric}
\def\ew{electroweak}
\def\xs{cross section}
\def\vev{vacuum expectation value}
\def\ep{$e^+e^-$}
\def\rg{renormalization group}
\def\lsp{lightest \susic\ particle}
\def\no{neutralino}
\def\co{chargino}
\def\sel{selectron}
\def\snu{sneutrino}
\def\eg{$e^-\gamma$}
\def\nwa{narrow width approximation}
\def\br{branching ratio}
\def\cm{centre of mass}

\def\gsim{\buildrel{\lower.7ex\hbox{$>$}}\over{\lower.7ex\hbox{$\sim$}}}
\def\lsim{\buildrel{\lower.7ex\hbox{$<$}}\over{\lower.7ex\hbox{$\sim$}}}
%
\catcode`@=11 
%
%
%

\font\fourteenrm=cmr10 scaled\magstep2
\font\twelverm=cmr10 scaled\magstep1
\font\ninerm=cmr9
\font\eightrm=cmr8
\font\sixrm=cmr6
\font\seventeenbf=cmbx10 scaled\magstep3
\font\fourteenbf=cmbx10 scaled\magstep2
\font\twelvebf=cmbx10 scaled\magstep1
\font\ninebf=cmbx9
\font\eightbf=cmbx8
\font\sixbf=cmbx6
\font\seventeeni=cmmi10 scaled\magstep3     \skewchar\seventeeni='177
\font\fourteeni=cmmi10 scaled\magstep2      \skewchar\fourteeni='177
\font\twelvei=cmmi10 scaled\magstep1        \skewchar\twelvei='177
\font\twelvei=cmmi10                        \skewchar\twelvei='177
\font\ninei=cmmi9                           \skewchar\ninei='177
\font\eighti=cmmi8                          \skewchar\eighti='177
\font\sixi=cmmi6                            \skewchar\sixi='177
\font\seventeensy=cmsy10 scaled\magstep3    \skewchar\seventeensy='60
\font\fourteensy=cmsy10 scaled\magstep2     \skewchar\fourteensy='60
\font\twelvesy=cmsy10 scaled\magstep1       \skewchar\twelvesy='60
\font\ninesy=cmsy9                          \skewchar\ninesy='60
\font\eightsy=cmsy8                         \skewchar\eightsy='60
\font\sixsy=cmsy6                           \skewchar\sixsy='60

\font\fourteenex=cmex10 scaled\magstep2
\font\twelveex=cmex10 scaled\magstep1
\font\twelveex=cmex10

\font\fourteensl=cmsl10 scaled\magstep2
\font\twelvesl=cmsl10 scaled\magstep1
\font\ninesl=cmsl9
\font\eightsl=cmsl8

\font\fourteenit=cmti10 scaled\magstep2
\font\twelveit=cmti10 scaled\magstep1
\font\nineit=cmti9
\font\eightit=cmti8
\font\twelvett=cmtt10 scaled\magstep1
\font\ninett=cmtt9
\font\eighttt=cmtt8
\font\twelvecp=cmcsc10 scaled\magstep1
\font\tencp=cmcsc10
\newfam\cpfam
%
%
%
%
\fontdimen16\seventeensy=4.59pt \fontdimen17\seventeensy=4.59pt
\fontdimen16\fourteensy=3.78pt \fontdimen17\fourteensy=3.78pt
\fontdimen16\twelvesy=3.24pt \fontdimen17\twelvesy=3.24pt
\fontdimen16\tensy=2.70pt \fontdimen17\tensy=2.70pt
\fontdimen16\ninesy=2.43pt \fontdimen17\ninesy=2.43pt
\fontdimen16\eightsy=2.16pt \fontdimen17\eightsy=2.16pt
\fontdimen16\sixsy=1.62pt \fontdimen17\sixsy=1.62pt
%
%
%
\newcount\f@ntkey\f@ntkey=0
\def\samef@nt{\relax \ifcase\f@ntkey \rm \or \oldstyle \or\or
  \or \it \or \sl \or \bf \or \tt \or \caps  \fi}
%
%
%
\def\eightpoint{\relax
  \textfont0=\eightrm \scriptfont0=\sixrm \scriptscriptfont0=\fiverm
  \textfont1=\eighti \scriptfont1=\sixi \scriptscriptfont1=\fivei
  \textfont2=\eightsy \scriptfont2=\sixsy \scriptscriptfont2=\fivesy
  \textfont3=\tenex \scriptfont3=\tenex \scriptscriptfont3=\tenex
  \textfont\itfam=\eightit
  \textfont\slfam=\eightsl
  \textfont\bffam=\eightbf \scriptfont\bffam=\sixbf
    \scriptscriptfont\bffam=\fivebf
  \textfont\ttfam=\eighttt
  \textfont\cpfam=\tencp
  \def\rm{\fam0 \eightrm \f@ntkey=0 }%
  \def\oldstyle{\fam1 \eighti \f@ntkey=1 }%
  \def\it{\fam\itfam \eightit \f@ntkey=4 }%
  \def\sl{\fam\slfam \eightsl \f@ntkey=5 }%
  \def\bf{\fam\bffam \eightbf \f@ntkey=6 }%
  \def\tt{\fam\ttfam \eighttt \f@ntkey=7 }%
  \def\caps{\fam\cpfam \tencp \f@ntkey=8 }%
  \h@big=6.8\p@{}\h@Big=9.2\p@{}\h@bigg=11.6\p@{}\h@Bigg=14\p@{}%
  \setbox\strutbox=\hbox{\vrule height 6.8pt depth 2.8pt width\z@}%
  \samef@nt }
\def\ninepoint{\relax
  \textfont0=\ninerm \scriptfont0=\sixrm \scriptscriptfont0=\fiverm
  \textfont1=\ninei \scriptfont1=\sixi \scriptscriptfont1=\fivei
  \textfont2=\ninesy \scriptfont2=\sixsy \scriptscriptfont2=\fivesy
  \textfont3=\tenex \scriptfont3=\tenex \scriptscriptfont3=\tenex
  \textfont\itfam=\nineit
  \textfont\slfam=\ninesl
  \textfont\bffam=\ninebf \scriptfont\bffam=\sixbf
    \scriptscriptfont\bffam=\fivebf
  \textfont\ttfam=\ninett
  \textfont\cpfam=\tencp
  \def\rm{\fam0 \ninerm \f@ntkey=0 }%
  \def\oldstyle{\fam1 \ninei \f@ntkey=1 }%
  \def\it{\fam\itfam \nineit \f@ntkey=4 }%
  \def\sl{\fam\slfam \ninesl \f@ntkey=5 }%
  \def\bf{\fam\bffam \ninebf \f@ntkey=6 }%
  \def\tt{\fam\ttfam \ninett \f@ntkey=7 }%
  \def\caps{\fam\cpfam \tencp \f@ntkey=8 }%
  \h@big=7.65\p@{}\h@Big=10.35\p@{}\h@bigg=13.05\p@{}\h@Bigg=15.75\p@{}%
  \setbox\strutbox=\hbox{\vrule height 7.65pt depth 3.15pt width\z@}%
  \samef@nt}
\def\tenpoint{\relax
  \textfont0=\tenrm \scriptfont0=\sevenrm \scriptscriptfont0=\fiverm
  \textfont1=\teni \scriptfont1=\seveni \scriptscriptfont1=\fivei
  \textfont2=\tensy \scriptfont2=\sevensy \scriptscriptfont2=\fivesy
  \textfont3=\tenex \scriptfont3=\tenex \scriptscriptfont3=\tenex
  \textfont\itfam=\tenit
  \textfont\slfam=\tensl
  \textfont\bffam=\tenbf \scriptfont\bffam=\sevenbf
    \scriptscriptfont\bffam=\fivebf
  \textfont\ttfam=\tentt
  \textfont\cpfam=\tencp
  \def\rm{\fam0 \tenrm \f@ntkey=0 }%
  \def\oldstyle{\fam1 \teni \f@ntkey=1 }%
  \def\it{\fam\itfam \tenit \f@ntkey=4 }%
  \def\sl{\fam\slfam \tensl \f@ntkey=5 }%
  \def\bf{\fam\bffam \tenbf \f@ntkey=6 }%
  \def\tt{\fam\ttfam \tentt \f@ntkey=7 }%
  \def\caps{\fam\cpfam \tencp \f@ntkey=8 }%
  \h@big=8.5\p@{}\h@Big=11.5\p@{}\h@bigg=14.5\p@{}\h@Bigg=17.5\p@{}%
  \setbox\strutbox=\hbox{\vrule height 8.5pt depth 3.5pt width\z@}%
  \samef@nt}
\def\twelvepoint{\relax
  \textfont0=\twelverm \scriptfont0=\ninerm \scriptscriptfont0=\sixrm
  \textfont1=\twelvei \scriptfont1=\ninei \scriptscriptfont1=\sixi
  \textfont2=\twelvesy \scriptfont2=\ninesy \scriptscriptfont2=\sixsy
  \textfont3=\twelveex \scriptfont3=\twelveex \scriptscriptfont3=\twelveex
  \textfont\itfam=\twelveit
  \textfont\slfam=\twelvesl \scriptfont\slfam=\ninesl
  \textfont\bffam=\twelvebf \scriptfont\bffam=\ninebf
    \scriptscriptfont\bffam=\sixbf
  \textfont\ttfam=\twelvett
  \textfont\cpfam=\twelvecp
  \def\rm{\fam0 \twelverm \f@ntkey=0 }%
  \def\oldstyle{\fam1 \twelvei \f@ntkey=1 }%
  \def\it{\fam\itfam \twelveit \f@ntkey=4 }%
  \def\sl{\fam\slfam \twelvesl \f@ntkey=5 }%
  \def\bf{\fam\bffam \twelvebf \f@ntkey=6 }%
  \def\tt{\fam\ttfam \twelvett \f@ntkey=7 }%
  \def\caps{\fam\cpfam \twelvecp \f@ntkey=8 }%
  \h@big=10.2\p@{}\h@Big=13.8\p@{}\h@bigg=17.4\p@{}\h@Bigg=21.0\p@{}%
  \setbox\strutbox=\hbox{\vrule height 10pt depth 4pt width\z@}%
  \samef@nt}
\def\fourteenpoint{\relax
  \textfont0=\fourteenrm \scriptfont0=\tenrm \scriptscriptfont0=\sevenrm
  \textfont1=\fourteeni \scriptfont1=\teni \scriptscriptfont1=\seveni
  \textfont2=\fourteensy \scriptfont2=\tensy \scriptscriptfont2=\sevensy
  \textfont3=\fourteenex \scriptfont3=\fourteenex
    \scriptscriptfont3=\fourteenex
  \textfont\itfam=\fourteenit
  \textfont\slfam=\fourteensl \scriptfont\slfam=\tensl
  \textfont\bffam=\fourteenbf \scriptfont\bffam=\tenbf
    \scriptscriptfont\bffam=\sevenbf
  \textfont\ttfam=\twelvett
  \textfont\cpfam=\twelvecp
  \def\rm{\fam0 \fourteenrm \f@ntkey=0 }%
  \def\oldstyle{\fam1 \fourteeni \f@ntkey=1 }%
  \def\it{\fam\itfam \fourteenit \f@ntkey=4 }%
  \def\sl{\fam\slfam \fourteensl \f@ntkey=5 }%
  \def\bf{\fam\bffam \fourteenbf \f@ntkey=6 }%
  \def\tt{\fam\ttfam \twelvett \f@ntkey=7 }%
  \def\caps{\fam\cpfam \twelvecp \f@ntkey=8 }%
  \h@big=11.9\p@{}\h@Big=16.1\p@{}\h@bigg=20.3\p@{}\h@Bigg=24.5\p@{}%
  \setbox\strutbox=\hbox{\vrule height 12pt depth 5pt width\z@}%
  \samef@nt }
%
%
%
\newdimen\h@big
\newdimen\h@Big
\newdimen\h@bigg
\newdimen\h@Bigg
\def\big#1{{\hbox{$\left#1\vbox to\h@big{}\right.\n@space$}}}
\def\Big#1{{\hbox{$\left#1\vbox to\h@Big{}\right.\n@space$}}}
\def\bigg#1{{\hbox{$\left#1\vbox to\h@bigg{}\right.\n@space$}}}
\def\Bigg#1{{\hbox{$\left#1\vbox to\h@Bigg{}\right.\n@space$}}}
%
%
%
\newskip\normaldisplayskip
\newskip\normaldispshortskip
\newskip\normalparskip
\newskip\normalskipamount
\normalbaselineskip = 20pt plus 0.2pt minus 0.1pt
\normallineskip = 1.5pt plus 0.1pt minus 0.1pt
\normallineskiplimit = 1.5pt
\normaldisplayskip = 20pt plus 5pt minus 10pt
\normaldispshortskip = 6pt plus 5pt
\normalparskip = 6pt plus 2pt minus 1pt
\normalskipamount = 5pt plus 2pt minus 1.5pt
%
%
\def\sp@cing#1{%
  \baselineskip=\normalbaselineskip%
    \multiply\baselineskip by #1%
    \divide\baselineskip by 12%
  \lineskip=\normallineskip%
    \multiply\lineskip by #1%
    \divide\lineskip by 12%
  \lineskiplimit=\normallineskiplimit%
    \multiply\lineskiplimit by #1%
    \divide\lineskiplimit by 12%
  \parskip=\normalparskip%
    \multiply\parskip by #1%
    \divide\parskip by 12%
  \abovedisplayskip=\normaldisplayskip%
    \multiply\abovedisplayskip by #1%
    \divide\abovedisplayskip by 12%
  \belowdisplayskip=\abovedisplayskip%
  \abovedisplayshortskip=\normaldispshortskip%
    \multiply\abovedisplayshortskip by #1%
    \divide\abovedisplayshortskip by 12%
  \belowdisplayshortskip=\abovedisplayshortskip%
    \advance\belowdisplayshortskip by \belowdisplayskip%
    \divide\belowdisplayshortskip by 2%
  \smallskipamount=\normalskipamount%
    \multiply\smallskipamount by #1%
    \divide\smallskipamount by 12%
  \medskipamount=\smallskipamount%
    \multiply\medskipamount by 2%
  \bigskipamount=\smallskipamount%
    \multiply\bigskipamount by 4 }
%
%
%
\newcount\fontsize
\def\Eightpoint{\eightpoint\fontsize=8\sp@cing{8}}
\def\Ninepoint{\ninepoint\fontsize=9\sp@cing{9}}
\def\Tenpoint{\tenpoint\fontsize=10\sp@cing{10}}
\def\Twelvepoint{\twelvepoint\fontsize=12\sp@cing{12}}
\def\Fourteenpoint{\fourteenpoint\fontsize=14\sp@cing{14}}
\newcount\spacesize
\def\singlespace{\spacesize=\fontsize
  \multiply\spacesize by 9\divide\spacesize by 12%
  \sp@cing{\spacesize}}
\def\normalspace{\sp@cing{\fontsize}}
\def\doublespace{\spacesize=\fontsize
  \multiply\spacesize by 15\divide\spacesize by 12%
  \sp@cing{\spacesize}}
\Twelvepoint 
\interlinepenalty=50
\interfootnotelinepenalty=5000
\predisplaypenalty=9000
\postdisplaypenalty=500
\hfuzz=1pt
\vfuzz=0.2pt
\catcode`@=12 
%

%
%
%
\newcount\chapterno\chapterno=0
\def\chapter#1{\global\advance\chapterno by 1\sectionno=0\vskip 0.2in
  \centerline{{\Fourteenpoint\bf\number\chapterno. #1}}
  \par\nobreak\vskip 0.1in
  \seteqnstr{\number\chapterno.}\equano=0
  \figno=0 }
%
%
\newcount\sectionno\sectionno=0
\def\section#1{\global\advance\sectionno by 1\vskip 0.2in
  \ifnum\chapterno=0
  \noindent{\bf\number\sectionno. #1}
  \else
  \noindent{\bf\eqnstr\number\sectionno. #1}
  \fi
  \par\nobreak\vskip 0.1in}
%
%
\def\appendix#1#2{\chapterno=-1\sectionno=0\vskip0.2in
  {\Fourteenpoint\bf\centerline{\bf Appendix #1.}\centerline{#2}}
  \par\nobreak\vskip 0.1in
  \seteqnstr{#1.}\equano=1 }
%
%
\def\seteqnstr#1{\relax\xdef\eqnstr{#1}}
\seteqnstr{}
\newcount\equano\equano=0
\def\eqname#1{\relax\global\advance\equano by 1
  \xdef#1{{\noexpand\rm(\eqnstr\number\equano)}}}
\def\eqn#1{\eqno\eqname{#1}#1}

\newcount\eqnlet
\def\eqaset#1{\eqnlet=96\global\advance\equano by 1\eqalignno{#1}}

\def\eqngrp#1{\global\advance\equano by -1\eqname{#1}}
\def\eqnameset#1{\relax\global\advance\eqnlet by 1
  \xdef#1{{\noexpand\rm(\eqnstr\number\equano\char\number\eqnlet)}}}
%
%
\def\backeq#1{\ifx#1\undefined\hbox{($\ast$)}%
  \message{*Eqn*}\else#1\fi}
%
%
\newcount\figno\figno=0
\def\fig#1#2#3{\relax\global\advance\figno by 1
  \xdef#1{{\eqnstr\number\figno}}
  \topinsert\Tenpoint\vbox{\vskip#3\leftskip2.0cm\rightskip2.0cm
  \noindent{{\bf Figure #1}\quad#2\par}}\endinsert}
\def\pagenumbers{\footline={\hss\tenrm\folio\hss}}
%
%
%
\newcount\footcount\footcount=0
\def\foot#1{\global\advance\footcount by 1
  \footnote{$^{\number\footcount}$}{#1}}
%
%
%
%
\newwrite\refout
\newcount\refno
\def\refbegin{\immediate\openout\refout=refout.tex\refno=1}

%
%
\def\form#1#2#3#4#5#6{\xdef#1{{\noexpand\rm#2, {\noexpand\it#3\/}
  {\noexpand\bf#4}, #5 (#6).}}}
%
%
\def\bookform#1#2#3#4{\xdef#1{{\noexpand\rm#2, in {\noexpand\it#3\/}\hfil
  \break\line{\hfill(#4).\qquad}}}}
%
%
%
%
\def\procform#1#2#3#4#5#6{\xdef#1{{\noexpand\rm#2, in {\noexpand\it#3\/},
  #4 (#5) #6.}}}
%
%
\def\prepform#1#2#3#4#5{\xdef#1{{\noexpand\rm#2, #3 (#4)#5.}}}
%
%
\def\freeform#1#2{\xdef#1{{\noexpand\rm#2.}}}
%
%
\def\refwrite#1{\immediate\write\refout{\noexpand#1}}
\def\ref#1#2{\ifx#1\undefined\message{*Ref*}$\ast$\else
  \ifx#2\undefined\xdef#2{\number\refno}#2%
  \refwrite{\item{#2.}\noexpand#1}%
  \global\advance\refno by 1\else#2\fi\fi}
%
%
\def\xref#1#2{\ifx#1\undefined\message{*Ref*}$\ast$\else
  \ifx#2\undefined\xdef#2{\number\refno}%
  \refwrite{\item{#2.}\noexpand#1}%
  \global\advance\refno by 1\else\fi\fi}
%
%
\def\sref#1{\ifx#1\undefined\message{*Ref*}$\ast$\else
  \refwrite{\item{}\noexpand#1}\fi}

\freeform{\art }{and references therein }
\freeform{\bmm}{A. Bartl, W. Majerotto and B. M\"o\ss lacher, these Lecture
Notes}
\form{\bfm}{A. Bartl, H. Fraas and W. Majerotto}{Z.\ Phys.}{C 34}{411}{1987}
\form{\bfmA}{A. Bartl, H. Fraas and W. Majerotto}{Nucl.\ Phys.}{B 297}{479
}{1988}
\form{\hk}{H.E. Haber and G.L. Kane}{Phys.\ Rep.}{117}{75}{1985}
\form{\hpn}{H.P. Nilles}{Phys.\ Rep.}{110}{1}{1984}
\form{\kr}{H. Komatsu and R. R\"uckl}{Nucl.\ Phys.}{B 299}{407}{1988}
\form{\ammr}{G. Altarelli, G. Martinelli, B. Mele and R. R\"uckl}{Nucl.\ Phys.
}{B 262}{204}{1985}
\form{\gkps}{I.F. Ginzburg, G.L. Kotkin, S.L. Panfil and V.G. Serbo}{
Nucl.\ Phys.}{B 228}{285}{1983}
\form{\gp}{J.A. Grifols and R. Pascual}{Phys.\ Lett.}{B 135}{319}{1984}
\form{\abf}{U. Amaldi, W. de Boer and H. F\"urstenau}{Phys.\ Lett.}{B
260}{447}{1991}
\form{\ghh}{M.K. Gaillard, L. Hall and I. Hinchliffe}{Phys.\ Lett.}{B
116}{279}{1982}
\form{\kl}{W.-Y. Keung and L. Littenberg}{Phys.\ Rev.}{D 28}{1067}{1983}
\form{\gkst}{I.F. Ginzburg, G.L. Kotkin, V.G. Serbo and V.I.
Telnov}{Nucl.\ Instr. Meth.}{205}{47}{1983}
\procform{\fp}{F. Pauss}{Large Hadron Collider Workshop}{eds G. Jarlskog
and D. Rein}{CERN 90-10, ECFA 90-133}{Vol.\ I, p.\ 118}
\procform{\bmmos}{A. Bartl et al.}{Large Hadron Collider Workshop}{eds G.
Jarlskog and D. Rein}{CERN 90-10, ECFA 90-133}{Vol.\ II, p.\ 1033}
\procform{\elpa}{J. Ellis and F. Pauss}{Workshop on Physics at Future
Accelerators}{ed. J.H. Mulvey}{CERN 87-07}{Vol.\ I, p.\ 80}
\prepform{\cor}{F. Cuypers, G.J. van Oldenborgh and R. R\"uckl}{Munich
preprint MPI-Ph/92/14}{1992}{}

\refbegin
\twelvepoint
\nopagenumbers
\singlespace
{\obeylines
\hfill January 1992
\hfill LMU/92/2
\hfill MPI-Ph/92/14
\hfill (REVISED)
\vfill
\centerline{\seventeenbf Supersymmetric Signals in Electron-Photon Collisions}
\vfill
\centerline{
Frank Cuypers$^a$\footnote{${}^*$}%
{\ninepoint Email: {\tt frank@hep.physik.uni-muenchen.de}}
,\
Geert Jan van Oldenborgh$^a$\footnote{${}^\dagger$}%
{\ninepoint Email: {\tt gj@hep.physik.uni-muenchen.de}}
and\
Reinhold R\"uckl$^{a,b}$\footnote{${}^\ddagger$}%
{\ninepoint Email: {\tt rer@dm0mpi11.bitnet}}
}
\vfill
\centerline{\it $^a$Sektion Physik, Universit\"at M\"unchen,
Theresienstra\ss e 37}
\centerline{\it D-8000 M\"unchen 2, Germany}
\bigskip
\centerline{\it $^b$Max-Planck-Institut f\"ur Physik,
Werner-Heisenberg-Institut, F\"ohringer Ring 6}
\centerline{\it D-8000 M\"unchen 40, Germany}
\vfill
\centerline{\bf ABSTRACT}
}

\doublespace

{\leftskip1.5cm\rightskip1.5cm
\noindent
Associated \sel-\no\ production in the process
$e^-\gamma\to\tilde e^-\tilde\chi^0$
provides a striking \susic\ signal:
events with a single high $p_\perp$ electron
and otherwise only invisible particles.
For $e^-\gamma$ collisions obtained at high energy linear colliders
through back-scattering of a laser beam,
this reaction is shown to be complementary to
\sel\ pair production
in the processes
$e^+e^-\to\tilde e^+\tilde e^-$
and
$e^-e^-\to\tilde e^-\tilde e^-$,
and to be a probe of heavy \sel s
beyond the kinematical limit of pair production.
The \sm\ background from
$e^-\gamma\to e^-Z^0$ and $W^-\nu$
is studied and substantially reduced
by rapidity and transverse momentum cuts.
The minimum required integrated luminosities for observing this \susic\ signal
are given as functions of several model parameters
and collider energies.
\par}
\vfill
\eject
\pagenumbers
\advance\pageno by -1


\chapter{Introduction}
Supersymmetry is often considered
the best motivated and most predictive candidate for new physics
at the TeV energy scale.
Recently,
the attractive features of \susy\ have again been underlined
by the observation that the minimal \susic\ extension of the \sm\
is consistent with grand unification and proton stability,
provided the symmetry breaking scale is not too high
[\ref\abf\ABF].
In contrast,
the plain \sm\ is not.
Of course,
this only provides a very indirect argument in favour of \susy,
but it encourages the search for direct evidence of the existence
of \susic\ partners of the \sm\ particles
at the next generation of high energy colliders.

Strongly interacting \susic\ particles
could be observed at the Large Hadron Collider (LHC) and
Superconducting Supercollider (SSC),
where the existence of squarks and gluinos can be probed
up to masses of the order of 1 TeV.
In comparison,
the potential in discovering sleptons, charginos and neutralinos
is quite limited
[\ref\fp\FP].
Ideal machines to search for \ew\ superpartners are
$e^+e^-$ colliders.
At least in the case of charged particles,
masses
$\tilde m\lsim\sqrt{s_{ee}}/2$
can be directly produced,
$\sqrt{s_{ee}}$ being the $e^+e^-$ \cm\ energy.
Thus,
in order to reach the TeV mass range in the \ew\ sector as well,
one needs an $e^+e^-$ linear collider such as
the CERN Linear Collider (CLIC)
[\ref\ep\EP].

However,
this does not necessarily mean that $e^+e^-$ colliders
in the energy range, say,
$\sqrt{s_{ee}}=500-1000$ GeV
are not competitive or uninteresting
{}from the point of view of \susy\ searches.
The point is that
\susic\ particles which only have \ew\ interactions
may well be lighter than the strongly interacting squarks and gluinos.
In fact,
such a pattern is expected in a minimal supergravity model
[\ref\hk\HK\sref\hpn]
where the
$SU(3)_c\otimes SU(2)_L\otimes U(1)_Y$
gaugino mass parameters
$M_3,\ M_2$ and $M_1$,
respectively,
are related to a single gaugino mass parameter $m_{1/2}$
by \rg\ equations.
Assuming
$M_3=M_2=M_1=m_{1/2}$
at the grand unification scale,
one finds at low energies
$M_3>M_2>M_1$.
The effective gluino mass is given by $M_3$,
while $M_2$ and $M_1$ enter the \co\ and \no\ mass matrices.
The model also provides \rg\ equations
for scalar masses involving essentially a gravitino mass parameter
$m_0$ and $M_{1,2,3}$.
Unless $m_0 \gg M_{1,2,3}$,
sleptons are predicted to be lighter than squarks,
$m_{\tilde l}<m_{\tilde q}$,
and partners of right-handed \sm\ fermions
tend to be somewhat lighter
than the partners of the corresponding left-handed ones,
$m_{\tilde f_R}\lsim m_{\tilde f_L}$.
As a side remark,
the $\tilde t$ squark plays a special role and may not fit in this pattern.
As a consequence,
an experimental bound on the \sel\ mass
puts similar constraints on the basic mass parameters of the model
as a considerably higher bound on the squark mass,
for example.
In this sense,
searches for sleptons at $e^+e^-$ colliders
in the few hundred GeV range
are competitive to squark searches at multi-TeV hadron colliders.

In this paper,
we study the prospects for producing and detecting
\sel s at linear $e^+e^-$ colliders
in the energy range $\sqrt{s_{ee}}\approx500-2000$ GeV{}.
More specifically,
we compare single $\tilde e^-$ production in $e^-\gamma$ collisions,
using Bremsstrahlung photons or back-scattered laser beams,
with $\tilde e^+\tilde e^-$ ($\tilde e^-\tilde e^-$) pair production
in $e^+e^-$ ($e^-e^-$) collisions
provided by the same linear collider.

\chapter{The Photon Beam}
Bremsstrahlung has a rather soft spectrum,
given by the familiar equivalent photon function
$$
P(y)=
{\alpha\over2\pi}
{1+(1-y)^2\over y}
\ln{s_{ee}\over m_e^2}
\ ,
\eqn\aa
$$
where $\alpha$ is the fine structure constant
and $y$ is the photon energy fraction
$E_\gamma/E_e$.
Since the energy of the effective $e^-\gamma$ collision
is substantially degraded
in comparison with the $e^+e^-$ collision energy,
this option is not expected to be an efficient way to produce heavy \sel s
[\ref\ghh\GHH].

A more energetic photon beam can be produced by back-scattering a high
intensity
laser ray on a high energy electron beam [\ref\gkst\GKST].
In principle,
the entire electron beam can be converted into photons.
These photons are then on-shell
and their spectrum is hard.
The distribution $P(y)$ of the energy fraction $y$ of the electron
transferred to a photon,
$y={E_{\gamma}/E_e}$,
is given by [\GKST]
$$
P(y)=
{1\over N}
\left(
1-y+{1\over1-y}-{4y\over x(1-y)}+{4y^2\over x^2(1-y)^2}
\right)
\ ,
\eqn\f
$$
where
$$
0\leq y\leq{x\over x+1}
\eqn\g
$$
and
$$
x={4E_eE_{laser}\over m_e^2}\ .
\eqn\h
$$
The factor $N$ normalizes $\int dy \,P(y)$ to 1.
The electron and laser beams are taken to be aligned
and their respective energies are
$E_e$ and $E_{laser}$.
In what follows,
we assume a 100\%\ conversion efficiency
and neglect the angular dispersion of the back-scattered photons.
We also choose
to take
$x=2(1+\sqrt{2})\approx4.83$ ,
which is the threshold for electron pair creation
in reactions of the back-scattered and laser photons
[\GKST].
For larger values of $x$
the pair creation process rapidly becomes so important that
the conversion efficiency drops considerably.
Nevertheless,
the effective $e^-\gamma$ energy is now comparable
to the $e^+e^-$ collision energy
and, equally important,
the $e^-\gamma$ luminosity is high,
to wit ${\cal L}_{e\gamma}\approx{\cal L}_{ee}\approx10^{33}$
cm$^{-2}$s$^{-1}$.
In the following,
results are presented for both types of photon beams.

\chapter{Signal and Backgrounds}
We concentrate on a particularly striking signal:
a single high $p_\perp$ electron.
This signal is obtained in minimal \susic\
extensions of the \sm\ [\HK]
where the \lsp\ (LSP) is the lightest \no\
$\tilde\chi^0_1$.
A \sel-\no\ pair
$\tilde e^-\tilde\chi^0_1$
is produced with the \sel\ subsequently decaying into an
electron-\no\ pair
$e^-\tilde\chi^0_1$.
This scenario assumes R-parity to be conserved,
so that the LSP is stable and remains unobserved.

One can also consider more complicated scenarios
involving the production and cascading decays of
\no s, \co s, \snu s and \sel s.
Provided at the end of such a shower we are left
with a single electron and only invisible
neutrinos and LSPs,
such a cascade mechanism yields a similar single electron signal,
with a less pronounced $p_\perp$ though.

Here we limit ourselves to the simplest process of this kind:
$$e^-\ \gamma\ \to\ \tilde e^-\ \tilde\chi^0_1\
\to\ e^-\ \tilde\chi^0_1\ \tilde\chi^0_1
\ .
\eqn\a
$$
Our calculations thus really provide a lower bound on the total \xs s
for producing the required signal from \susy\ at \eg\ colliders.
In that sense,
our results are conservative.
The \sm\ background to the signal arises from the following two reactions:
$$
\eqalign{
e^-\ \gamma\
&\to\ e^-\ Z^0\
\to\ e^-\ \bar\nu\ \nu\ ,\cr
e^-\ \gamma\
&\to\ W^-\ \nu\
\to\ e^-\ \bar\nu\ \nu\ .\cr
}
\eqn\b
$$
In this feasibility study
we only consider tree level processes in the \nwa.
So,
the $\tilde e^-$, $Z^0$ and $W^-$ are produced on shell
and are left to decay into respectively
$e^-\tilde\chi^0_1$, $\bar\nu\nu$ and $\bar\nu e^-$
with the corresponding \br s.

The results depend on four \susy\ parameters [\HK]:
\item{$\bullet$} the ratio
$\tan\theta_v=v_2/v_1$
of the Higgs \vev s;
\item{$\bullet$} the soft \susy\ breaking mass parameters
$M_2$ and $\mu$
associated with the $SU(2)_L$ gauginos and higgsinos, respectively
(for the $U(1)_Y$ gaugino mass parameter $M_1$ we assume
$M_1=M_25/3\tan^2\theta_w$, where $\theta_w$ is the weak mixing angle,
in accordance with the renormalization group evolution
{}from a common value $M_1=M_2$ at the GUT scale);
\item{$\bullet$} the mass of the \sel\ $m_{\tilde e}$
(for simplicity we assume the \susic\ partners
of the left- and right-handed electrons to have equal masses:
$m_{\tilde e_L}=m_{\tilde e_R}$).

For arbitrary values of $M_2$ and $\mu$
the gauginos and higgsinos mix to form \no\ and \co\ mass eigenstates
[\HK].
Roughly,
for $|\mu|\gsim M_2/2$
the higgsino admixture to the lightest \no\ is small
[\ref\kr\KR].
This is an essential condition to obtain measurable \xs s
for the process under study
since the higgsino-electron Yukawa coupling is suppressed
by the mass of the electron.
In this region of parameter space
the dependence of
$\sigma(e^-\gamma\to\tilde e^-\tilde\chi^0_1)$ on
$\theta_v$ remains very small.
If, in addition,
$M_2\gsim m_Z$,
one finds roughly
$m_{\tilde\chi^0_1}\approx{M_2/2}$
and
$m_{\tilde\chi^0_2}\approx m_{\tilde \chi^\pm_1}\approx M_2$
[\KR].
For consistency with the assumption that $\tilde\chi^0_1$ is the LSP
we must in this case require $M_2\lsim 2m_{\tilde e}$ .
To get a feeling for the boundaries in parameter space
which can be probed,
the reader may refer to Fig.\ 6,
where the parameter regions yielding favourable \xs s
are clearly displayed.

In what follows
we have considered four
different scenarios,
which are summarized in Table 1.
For the first scenario we have chosen
$\tan\theta_v=200$,
whereas for the three others we have chosen
$\tan\theta_v=2$.
The similarity between the output parameters of scenario 0 and scenario 1
illustrates that
$\theta_v$
is not an essential parameter
(as long as
$|\mu|\gsim M_2/2$).

The lowest order Feynman diagrams leading to the
$\tilde e^-\tilde\chi^0_1$,
$e^-Z^0$ and
$\nu W^-$
final states in \eg\ collisions
are shown in Figs 1.
The matrix elements and differential \xs s
$d\sigma/dt$
derived from these diagrams
are given in Ref.\ \ref\ammr\AMMR.
For the total \xs s one finds
(see also Refs \ref\gkps\GKPS\ and \ref\gp\GP)
$$
\eqalign{
\sigma
&
\left(e^-_{L,R}\gamma \to \tilde e^-_{L,R}\tilde\chi^0_1\right)
=
{\pi\alpha^2\over s^3}
{|G_{L,R}|^2\over2}
\cr
&\left[
(s-7m_{\tilde\chi^0_1}^2+7m_{\tilde e}^2)\sqrt{\lambda}
-
4(m_{\tilde e}^2-m_{\tilde\chi^0_1}^2)
(s+m_{\tilde e}^2-m_{\tilde\chi^0_1}^2)
\ln\left({s-m_{\tilde\chi^0_1}^2+m_{\tilde e}^2+\sqrt{\lambda}
\over s-m_{\tilde\chi^0_1}^2+m_{\tilde e}^2-\sqrt{\lambda}}\right)
\right]
\ ,
}
\eqn\ca
$$
$$
\eqalign{
\sigma
\left(e^-\gamma \to e^-Z^0\right)
&=
{\pi\alpha^2\over s^3}
{(1-4\sin^2\theta_w)^2+1\over4\sin^22\theta_w}\cr
&\left[
(s-m_Z^2)(s+3m_Z^2)
+
2(s^2-2sm_Z^2+2m_Z^4)
\ln\left({(s-m_Z^2)^2\over m_e^2s}\right)
\right]
\ ,
}
\eqn\cb
$$
$$
\eqalign{
\sigma
\left(e^-\gamma \to \nu W^-\right)
&=
{\pi\alpha^2\over s^3}
{1\over4\sin^2\theta_w}\cr
&\left[
{(s-m_W^2)(4s^2+5sm_W^2+7m_W^4)\over m_W^2}
-
4(2s^2+sm_W^2+m_W^4)\ln\left({s\over m_W^2}\right)
\right]
\ ,\cr
}
\eqn\cc
$$
where
$$
\lambda =
\lambda(s,m_{\tilde e}^2,m_{\tilde\chi^0_1}^2) =
s^2+m_{\tilde e}^4+m_{\tilde\chi^0_1}^4
-2sm_{\tilde e}^2-2m_{\tilde e}^2m_{\tilde\chi^0_1}^2-2m_{\tilde\chi^0_1}^2s
\ .
\eqn\d
$$
The
$e_L \tilde e_L \tilde\chi^0_1$
and
$e_R \tilde e_R \tilde\chi^0_1$
couplings
$G_L$ and $G_R$
depend on the photino
($\tilde \gamma$)
and zino
($\tilde Z$)
content of the lightest \no\
$\tilde\chi^0_1$
(we can safely ignore its higgsino
($\tilde H$)
admixture,
since the
$e \tilde e \tilde H$
coupling is proportional to the mass of the electron):
$$
\eqalign{
G_L&=
U_{11}Q + U_{21}{T_3-Q\sin^2\theta_w\over\sin\theta_w\cos\theta_w}\cr
G_R&=
U_{11}^*Q + U_{21}^*{-Q\sin\theta_w\over\cos\theta_w}\ ,\cr
}
\eqn\e
$$
where $Q=-1$ and $T_3=-1/2$
are the electrons charge and third component of the weak isospin
and $\theta_w$ is the weak mixing angle.
$U_{11}$ and $U_{21}$
are elements of the unitary matrix which diagonalizes the \no\ mass matrix
[\HK,\KR].
They, as well as the masses of the \no s,
depend in a non-trivial manner on
$\theta_v$, $\mu$ and $M_2$.

In the \xs\ formulas Eqs \ca-\cc\
we have neglected the mass of the electron $m_e$ everywhere,
except in Eq. \cb\
where $m_e$ has been kept in the electron propagator
of the second diagram of Fig.\ 1b.
Indeed,
the $u$-channel pole is only regularized by the finite mass of the electron.
As a result,
the electron distribution
of the $e^-\gamma \to e^-Z^0$ channel
is very strongly peaked in the photon direction.

Using Eqs \aa\ and \f\ we fold
all \xs s with the energy distribution $P(y)$
of the photon beam.
The laboratory frame is thus not the \cm\ frame
and the \cm\ energy $\sqrt{s_{e\gamma}}$ is given by
$$
s_{e\gamma}=ys_{ee}\ ,
\eqn\i
$$
where $\sqrt{s_{ee}}=2E_e$ is the collider energy.
The convoluted \xs s are obtained from
$$
\sigma(s_{ee}) =
\int_{y_{min}}^{y_{max}} dy\ P(y)\sigma(s_{e\gamma})
\ ,
\eqn\j
$$
where the upper integration limit $y_{max}$ is given
by 1 in the case of Bremsstrahlung
and by Eq. \g\ in the case of a back-scattered laser beam.
The lower limit $y_{min}$ is set by the kinematical threshold,
$y_{min}=(m_{\tilde e^-}+m_{\tilde\chi^0_1})^2/s_{ee}$
for the \susic\ process and
$y_{min}=m^2_{Z,W}/s_{ee}$
for the background processes.

In the \nwa, the \xs s for producing a single electron signal
are obtained by multiplying the folded \xs s Eq. \j\
by the appropriate \br s.
We have taken the
$Z^0\to\bar\nu\nu$
and
$W^-\to e^-\bar\nu$
\br s to be 20\% and 10\% respectively.
For the
$\tilde e^-\to e^-\tilde\chi^0_1$
decay [\ref\bfmA\BFMA],
the \br\ is 100\% if the \sel\ is lighter
than the second lightest \co\ or \no.
If there are \co s or other \no s
which are lighter than the \sel,
the
$\tilde e^-_L\to e^-_L\tilde\chi^0_1$
\br\ can be considerably less than 100\%.
As long as $|\mu|,M_2\gsim m_Z$, however,
the
$\tilde e^-_R\to e^-_R\tilde\chi^0_1$
\br\ remains close to 100\%.

\chapter{Results}
In Figs 2 total \xs s are plotted as functions of the \sel\ mass.
We compare the
$\tilde e^+\tilde e^-$
and
$\tilde e^-\tilde e^-$
production rates at $e^+e^-$ [\ref\bfm\BFM] and $e^-e^-$ [\ref\kl\KL]
colliders\foot{\ninepoint We have computed these \xs s
in the approximation where
$m_{\tilde\chi^0_1}\ll m_{\tilde\chi^0_i}$ ($i$=2--4).}
with the
$\tilde e^-\tilde\chi^0_1$
production at the same colliders operating in the $e^-\gamma$ mode.
This is done for scenarios 1-3
summarized in Table 1
and correspondingly three different collider energies:
$\sqrt{s_{ee}}=500,\ 1000,\ 2000$ GeV .
Although the $e^-\tilde\chi^0_1\tilde\chi^0_1$ channel \xs s
are lower for small \sel\ masses,
only this channel remains
when $m_{\tilde e}>\sqrt{s_{ee}}/2$.
Note that the scenarios considered here
yield relatively heavy \no s.
With lighter \no s the \xs s
extend even further beyond the kinematical limit of pair production.
We also plotted the \xs s obtained with Bremsstrahlung photons
off the electron beam in the Weizs\"acker-Williams approximation [\GHH].
As expected,
they are negligible in comparison with the \xs s obtained with
back-scattered laser photon beams.

Fig.\ 3 displays the behaviour of the total
$e^-\gamma\to e^-\tilde\chi^0_1\tilde\chi^0_1$
\xs\ as a function of the collider energy.
The four scenarios summarized in Table 1
are considered here,
and the \sel\ mass is set equal to $M_2$.
For scenario 0 the
$\tilde e^-_L\to e^-_L\tilde\chi^0_1$
\br\ is only 75.8\%
because the \sel\ can also decay into a \co\ and neutrino.
For scenarios 1-3, however,
the \sel\ can only decay into the LSP and electron.
One observes that scenarios 0 and 1,
which differ only by the value of $\theta_v$,
yield very similar results.
The \sm\
$e^-\gamma\to e^-\bar\nu\nu$
backgrounds are also shown.
It appears that the \susic\ signal is completely swamped.
However,
since the largest portion of the \sm\ \xs s
is due to the u-channel exchange in
$e^-\gamma\to e^-Z^0$
and the t-channel exchange in
$e^-\gamma\to W^-\nu$
(since $m_e,m_W\ll\sqrt{s_{ee}}$),
these \xs s are drastically reduced by an angular or rapidity cut.

The final electron
transverse momentum and rapidity distributions
are shown in Figs 4,
for the $Z^0$ and $W^-$ channels
and the $\tilde e^-$ channel in scenario 1.
The \sel\ mass and the electron beam energy have been chosen to be equal:
$m_{\tilde e}=\sqrt{s_{ee}}/2=250$ GeV{}.
One sees that
in the $Z^0$ and (to a lesser extent) $W^-$ channels
the final electron is preferentially emitted
in the direction of the incoming photon.
In contrast,
the $\tilde e^-$ channel displays very little preference,
the decay electron being produced centrally.

In order to enhance the \susic\ signal relative to the \sm\ background,
we have chosen to impose the following cuts:
$$\eqalign{
p_\perp>\sqrt{s_{ee}}/10\, \cr
0<\eta<2\ .
}
\eqn\k
$$
The cut \xs s are displayed in Figs 5
as functions of the \sel\ mass.
This is done for scenarios 1 and 2
of Table 1
and respectively two different collider energies:
$\sqrt{s_{ee}}=500,\ 1000$ GeV{}.
With these two cuts the backgrounds drop by more than one order of magnitude,
whereas for \sel\ masses higher than the beam energy
the \susic\ signal is reduced by about a factor 2.
For \sel\ masses much lower than the beam energy,
these cuts also dwarf the \susic\ signal,
but then anyway \sel\ pair production
in $e^+e^-$ and $e^-e^-$ collisions
is another promising mechanism
as is shown in Figs 2.

The \sm\ background can be computed with great precision.
The accuracy of these calculations can even be checked
in the case of the $W^-$ channel,
by comparing with the
$e^-\gamma\to \mu^-\bar\nu\nu$
signal.
In principle, thus,
any deviation from these predicted results
can be a signal for \susy.
Finite statistics,
however,
preclude too small a \susy\ signal
to emerge from the background statistical fluctuations.
Assuming Poisson statistics,
to obtain a
$R\sigma$
confidence level,
the number of signal events
$n_{\rm SUSY}$
needs to be larger than $R$ times
the square root of the number of background events
$n_{\rm SM}$:
$n_{\rm SUSY} \geq R\sqrt{n_{\rm SM}}$.
The luminosity required to separate
the \susic\ signal from its \sm\ background
is thus
$$
{\cal L}_{req}\geq{R^2\sigma_{\rm SM}\over\sigma_{\rm SUSY}^2}
\ .
\eqn\l
$$
Setting $R=3$
($3\sigma$ confidence level)
we have plotted on the right vertical axis of Figs 5
the minimum required integrated luminosities
for having a discovery potential.
Also,
specializing on a 250 GeV/beam machine
we have plotted in Fig.\ 6,
in the $(\mu,M_2)$-plane,
the contours of minimum required luminosity
for extracting the \susic\ signal from the background.
This plot has been produced
with $\tan\theta_v=2$
and for a \sel\ with a mass equal to the electron beam energy,
so that the \sel\ cannot be pair produced.
For values of $M_2$ larger than about 380 GeV
(for large absolute values of $\mu$)
the studied reaction is kinematically forbidden
since $m_{\tilde e} + m_{\tilde\chi_1^0} > \sqrt{s_{ee}}$.
The region where
$|\mu|\lsim M_2/2$
is inaccessible because of the large Higgsino content of the LSP.
The parameter region to be explored
by the Large Electron Positron Collider (LEP 200)
is also shown
[\FP,\ref\bmmos\BMMOS].

\chapter{Conclusions}
It appears possible to convert a high energy electron beam
into an almost as energetic and very intense photon beam
by back-scattering a suitable laser beam.
Linear colliders will thus eventually provide high energy
$e\gamma$ and $\gamma\gamma$ collisions,
in addition to $ee$ collisions.
We have studied to what extent the $e^-\gamma$ option
can be exploited for \susy\ searches.
For this purpose,
we have concentrated on the most interesting channel,
$e^-\gamma\to\tilde e^-\tilde\chi^0_1\to e^-\tilde\chi^0_1\tilde\chi^0_1$,
and compared its prospects to the expectations for
$\tilde e^+\tilde e^-$ and $\tilde e^-\tilde e^-$ pair production
in $e^+e^-$ and $e^-e^-$ collisions.
For the present study,
we assumed the lightest \no\ $\tilde\chi^0_1$
to be the \lsp\
and, therefore, stable and undetectable.

In summary,
our results indicate that the $e^-\gamma$ mode
is indeed a valuable option.
Not only does the
$e^-\tilde\chi^0_1\tilde\chi^0_1$
channel exhibit an extremely simple and striking signature,
but also the most dangerous background from the \sm\ processes
$e^-\gamma\to e^-Z^0\to e^-\bar\nu\nu$
and
$e^-\gamma\to W^-\nu\to e^-\bar\nu\nu$
is manageable.
Firstly,
this background can be calculated with great precision
because it only involves leptons and weak gauge bosons.
Secondly,
it can be substantially reduced experimentally by simple cuts
on the rapidity and transverse momentum of the final state electron.

In comparison to \sel\ searches
in $e^+e^-$ and  $e^-e^-$ collisions at the same linear collider,
the situation is as follows.
For
$m_{\tilde e}<\sqrt{s_{ee}}/2$,
$\tilde e^+\tilde e^-$ and $\tilde e^-\tilde e^-$
pair production
is more favourable than single $\tilde e^-$ production
in $e^-\gamma$ collisions
because of a larger signal \xs,
while the background problems are similar.
Nevertheless,
also in this lower mass range,
the $e^-\gamma$ mode is of considerable interest
as a cross-check of a signal which might be observed
in the $e^+e^-$ and  $e^-e^-$ collisions.
Moreover,
$e^-\gamma$ collisions provide a more direct probe of \no s
and their properties
than $e^+e^-$ collisions,
where one has to consider higher order processes
such as $e^+e^-\to\tilde\chi^0_1\tilde\chi^0_1\gamma$.

For
$m_{\tilde e}>\sqrt{s_{ee}}/2$,
the role of $e^-\gamma$ and $e^+e^-$ or $e^-e^-$ collisions
is  reversed.
In the $e^-\gamma$ mode the \sel\ can still be singly produced
and studied directly.
In contrast,
in the $e^+e^-$ and  $e^-e^-$ modes
it can only be probed indirectly via virtual effects
in the $e^+e^-\to\tilde\chi^0_1\tilde\chi^0_1\gamma$ channel.
{}From Figs 2 and 5 it appears, however,
that for a fixed integrated luminosity,
say $\int{\cal L}~dt = 20$ fb$^{-1}$,
this advantage can only be exploited experimentally
at collider energies $\sqrt{s_{ee}} < 1$ TeV.

\bigskip
We are grateful to Witold Kozanecki for sharing his expertise
on back-scattered photon beams
and to Walter Majerotto for useful advice.

\vfill
\eject
\immediate\closeout\refout
  \vskip0.2in\noindent{\bf References}\vskip0.2in
  {\obeylines\input refout }

\vfill
\eject
\parindent2cm

\centerline{Table Caption}

\bigskip
\item{Table 1: } Four typical scenarios parametrized in terms of
$\tan\theta_v,\mu$ and $M_2$,
and the resulting lightest \no\ and \co\ masses.
The unitary matrix elements $U_{11}\ (U_{21})$
give the photino (zino) content of the lightest \no.
The $\tilde e^-_L$ \br\
is given for
$m_{\tilde e}=M_2$,
while
$BR(\tilde e^-_R\to e^-_R\tilde\chi^0_1)\approx1$
for all cases.

\bigskip
\bigskip
\bigskip
\bigskip
\bigskip
\bigskip
\moveright1.5cm
\vbox{
\offinterlineskip
\halign{
\strut \vrule\vrule \hfil\hskip.1cm#\hskip.1cm\hfil & \vrule\hskip1pt\vrule
\hfil\hskip.1cm#\hskip.1cm\hfil & \vrule \hfil\hskip.1cm#\hskip.1cm\hfil &
\vrule \hfil\hskip.1cm#\hskip.1cm\hfil & \vrule\hskip1pt\vrule
\hfil\hskip.1cm#\hskip.1cm\hfil & \vrule \hfil\hskip.1cm#\hskip.1cm\hfil &
\vrule \hfil\hskip.1cm#\hskip.1cm\hfil & \vrule
\hfil\hskip.1cm#\hskip.1cm\hfil & \vrule \hfil\hskip.1cm#\hskip.1cm\hfil
\vrule\vrule \cr
\noalign{\hrule}
\noalign{\hrule}
sce- & $\tan\theta_v$ & $\mu$ & $M_2$ & $m_{\tilde\chi_1^0}$ &
$m_{\tilde\chi_1^-}$ & $U_{11}$ & $U_{21}$ & BR \cr
nario &  & [GeV] & [GeV] & [GeV] & [GeV] &  &  &
$\tilde e_L^-\to e_L^- \tilde\chi_1^0$ \cr
\noalign{\hrule}
\noalign{\hrule}
0 & 200 & -375 & 250 & 123 & 233 & .855 & -.498 & .76  \cr
\noalign{\hrule}
1 & 2 & -375 & 250 & 127 & 255 & .891 & -.445 & 1 \cr
\noalign{\hrule}
2 & 2 & -750 & 500 & 250 & 502 & .881 & -.471 & 1 \cr
\noalign{\hrule}
\noalign{\hrule}
}
\vskip1cm
\moveright-1.5cm
\centerline{Table 1}
}

\vfill
\eject

\centerline{Figure Captions}

\bigskip
\item{Figs 1: } Lowest order Feynman diagrams of the processes
\item{} (a) $e^-\gamma \to \tilde e^-\tilde\chi^0_1$,
\item{} (b) $e^-\gamma \to e^-Z^0$,
\item{} (c) $e^-\gamma \to W^-\nu$.

\bigskip
\item{Figs 2: } Total \xs s as functions of the \sel\ mass for
$e^-\gamma \to \tilde e^-\tilde\chi^0_1 \to e^-\tilde\chi^0_1\tilde\chi^0_1$
convoluted with the energy spectrum of back-scattered laser photons
(full curves),
$e^-\gamma \to \tilde e^-\tilde\chi^0_1 \to e^-\tilde\chi^0_1\tilde\chi^0_1$
convoluted with a Bremsstrahlung spectrum
(dotted curves),
$e^-e^- \to \tilde e^-\tilde e^-$
(dot-dashed curves) and
$e^+e^- \to \tilde e^+\tilde e^-$
(dashed curves).
These \xs s are shown for three different collider energies
and \susic\ scenarios:
\item{} (a) $\sqrt{s_{ee}}=500$ GeV and scenario 1;
\item{} (b) $\sqrt{s_{ee}}=1000$ GeV and scenario 2;
\item{} (c) $\sqrt{s_{ee}}=2000$ GeV and scenario 3.

\bigskip
\item{Fig.\ 3: } Signal and background total \xs s
as functions of the collider energy $\sqrt{s_{ee}}$ for
$e^-\gamma \to \tilde e^-\tilde\chi^0_1 \to e^-\tilde\chi^0_1\tilde\chi^0_1$
and scenarios 1-3
(full curves) and 0
(dotted curve)
with $m_{\tilde e}=M_2$,
$e^-\gamma \to e^-Z^0 \to e^-\bar\nu\nu$
(dot-dashed curve) and
$e^-\gamma \to W^-\nu \to e^-\bar\nu\nu$
(dashed curve).

\vfill
\eject
\bigskip
\item{Figs 4: }
Transverse momentum (a) and rapidity (b) distributions
of the final electron in the processes:
$e^-\gamma \to \tilde e^-\tilde\chi^0_1 \to e^-\tilde\chi^0_1\tilde\chi^0_1$
for scenario 1 and $m_{\tilde e}=250$ GeV
(diamonds),
$e^-\gamma \to e^-Z^0 \to e^-\bar\nu\nu$
(crosses) and
$e^-\gamma \to W^-\nu \to e^-\bar\nu\nu$
(squares),
at $\sqrt{s_{ee}}=500$ GeV{}.

\bigskip
\item{Figs 5: } Signal and background \xs s
after kinematical cuts
as functions of the \sel\ mass for
$e^-\gamma \to \tilde e^-\tilde\chi^0_1 \to e^-\tilde\chi^0_1\tilde\chi^0_1$
(full curves),
$e^-\gamma \to e^-Z^0 \to e^-\bar\nu\nu$
(dotted lines) and
$e^-\gamma \to W^-\nu \to e^-\bar\nu\nu$
(dot-dashed lines).
These \xs s are shown for two different collider energies
and \susic\ scenarios:
\item{} (a) $\sqrt{s_{ee}}=500$ GeV and scenario 1;
\item{} (b) $\sqrt{s_{ee}}=1000$ GeV and scenario 2;
\item{} The luminosities required for extracting the \susic\ signal
{}from the background at a $3\sigma$ confidence level
are shown on the right vertical axis.

\bigskip
\item{Fig.\ 6: } Contours in the ($\mu,M_2$)-plane
showing the minimum required integrated luminosities
for obtaining a $3\sigma$ effect
for $\sqrt{s_{ee}}=500$ GeV,
$m_{\tilde e}=250$ GeV
and $\tan\theta_v=2$.
The dark grey areas are excluded
since there the \sel\ would be lighter than the lightest \no.
The light grey region will be explored by LEP 200
[\FP,\BMMOS].

\end